# Nonreciprocal Elasticity

## Mohamed Shaat

*Mechanical Engineering Department, Abu Dhabi University, Al Ain, P.O.BOX 1790, United Arab Emirates.*

Email: mohamed.i@adu.ac.ae & shaatscience@yahoo.com

**Nonreciprocity has been introduced to various fields to realize asymmetric, nonlinear, and/or time non-revisal physical systems. By virtue of the Maxwell-Betti reciprocal theorem, breaking the time-reversal symmetry of dynamic mechanical systems is only possible using nonlinear materials. Nonetheless, nonlinear materials should be accompanied by geometrical asymmetries to achieve nonreciprocity in static systems. Here, we further investigate this and demonstrate a novel "nonreciprocal elasticity" concept. We show that the nonreciprocity of static mechanical systems can be achieved only and only if the material exhibits nonreciprocal elasticity. We experimentally demonstrate linear and nonlinear materials with nonreciprocal elasticities. By means of topological mechanics, we demonstrate that the mechanical nonreciprocity requires nonreciprocal elasticity no matter what the material is linear or nonlinear elastic. We show that linear materials with nonreciprocal elasticity can realize nonreciprocal-topological systems. The nonreciprocal elasticity developed here will open new venues of the design of mechanical systems with effective nonreciprocity.**

## Introduction

Nonreciprocity has been introduced to various fields to give asymmetrical, nonlinear, and/or time non-revisal physical systems[1–9]. Optical nonreciprocity has been recently introduced to photonics, optical diodes, and insulators to give nonreciprocal transmissions of light fields[3–7]. In addition, nonreciprocity was introduced to realize mechanical systems with topological characteristics, e.g., nonreciprocal waves[1], static nonreciprocity[2,10], and nonreciprocal edge states[8]. In many occasions, the optical nonreciprocity was achieved using nonlinear optical systems, which disobey the Lorentz reciprocity law[6]. Similar to optical systems, achieving the nonreciprocity in mechanical systems requires breaking the symmetry of the material deformation using nonlinear materials, which disobey the Maxwell-Betti reciprocal theorem[1,11–16]. Nonetheless, materials or structures that would achieve asymmetric couplings of certain fields would induce nonreciprocity without breaking the reciprocity laws. This has been realized in optical systems where structures that achieved asymmetric couplings of the optical fields induced optical nonreciprocity without breaking the Lorentz reciprocity law[9].

Achieving the nonreciprocity in mechanical systems requires breaking the symmetry of the material deformation. By virtue of the Maxwell-Betti reciprocal theorem, nonlinear materials can impart some nonreciprocal characteristics to mechanical systems. Therefore, nonlinear materials have been utilized to realize nonreciprocal mechanical systems, especially the dynamical ones. However, for static mechanical systems, using nonlinear materials is not enough to realize effective nonreciprocity[2]. It was advised that combining large nonlinearities and microstructural geometrical asymmetries would achieve strong static nonrecirpcoity[2]. Here, we further investigate this and show that the static mechanical nonreciprocity can be achieved only and only if the material exhibits "nonreciprocal elasticity". We experimentally demonstrate an example of a metamaterial with nonreciprocal elasticity. By means of topological mechanics, we demonstrate that the mechanical nonreciprocity requires nonreciprocal elasticity no matter what the material is linear or nonlinear elastic. We demonstrate that not only nonlinear materials but also linear materials can achieve asymmetric deformation and mechanical nonreciprocity.

**Results**

Consider an elastic homogeneous material that is setup under a tensile test such that it is fixed from its side B and stretched from its side A ($A \rightarrow B$), as shown in Fig. 1(a). This material is expected to exhibit the same elastic properties even if we switched its setup such that it became fixed from side A and stretched from side B ($B \rightarrow A$) (Fig. 1(a)). This is expected, as we never considered the direction of the material sample with respect to the applied force direction during the tensile test. A material with this behavior is a "*reciprocal elastic*" material. Reciprocal elasticity is a common property of all natural materials including both linear and nonlinear materials. When either linear or nonlinear material is stretched by the same tensile force from either side A or side B, the material shows the same stiffness, and the measured displacement is the same.

Here, we demonstrate a metamaterial that exhibits "*nonreciprocal elasticity*". Nonreciprocal elasticity can be defined as the contrast in the material's elastic properties along two opposite directions. In other words, a nonreciprocal elastic material is the one that exhibits asymmetric elastic properties, which are different when measured along two opposite directions. The developed metamaterial was designed with a special periodic microstructure to achieve nonreciprocal elasticity (Fig. 1(b)). This periodic microstructure is featured with a geometrical angle $\theta$ (see Fig. 1(b)). This angle $\theta$ gave control to the degree of the nonreciprocal elasticity achieved by the metamaterial. Six models of the metamaterial with different geometrical angles, $\theta = 0°, 15°, 30°, 45°, 60°$, and $75°$ were 3D-printed and tested under tension. Each material sample was tested under the two tensile setups shown in Fig. 1(a), i.e., $A \rightarrow B$ and $B \rightarrow A$. The experimental results of the tensile tests of the various material samples are represented in Figs. 1(c) and 1(d).

Tailoring the geometrical angle, $\theta$, gave us the ability to tailor the degree of the material's nonreciprocal elasticity. We defined, $\epsilon$, as a nonreciprocal elasticity measure:

$$\epsilon = 1 - \frac{E_{B \rightarrow A}}{E_{A \rightarrow B}} \text{ or } 1 - \frac{E_{A \rightarrow B}}{E_{B \rightarrow A}} \qquad (1)$$

where $E_{A \rightarrow B}$ ($E_{B \rightarrow A}$) is the material's elastic modulus measured by the tensile test such that it is fixed from its side B (A) and stretched from its side A (B). When $\epsilon = 0$, the material is reciprocal elastic and $E_{B \rightarrow A} = E_{A \rightarrow B}$. When $\epsilon \neq 0$, the material exhibits nonreciprocal elasticity. A typical range of the nonreciprocal elasticity parameter is $-\infty < \epsilon < 1$. Fig. 1(c) shows the variation of the nonreciprocal elasticity parameter, $\epsilon$, as a function of the geometrical angle, $\theta$.

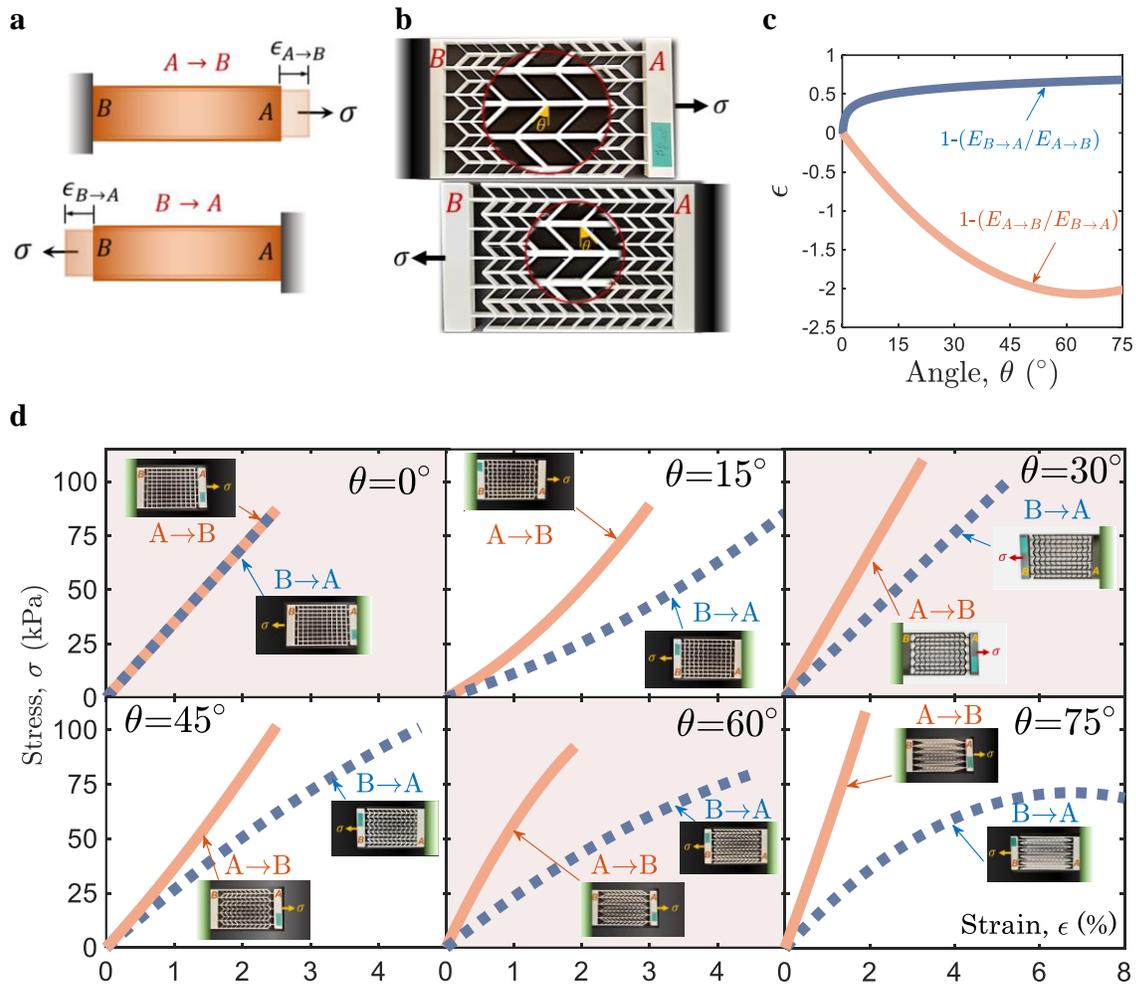

**Fig. 1: Metamaterials with Nonreciprocal Linear and Nonlinear Elasticities. a,** An elastic material under two different setups of tensile testing. In the setup $A \to B$ ($A \to B$), the material is fixed at side B (A) and stretched from side A (B). **b,** Two samples of the developed metamaterial with a geometrical asymmetric angle $\theta = 45°$. **c,** The nonreciprocal elasticity parameter $\epsilon$ versus the geometrical angle $\theta$ (experimental results). **d,** The elastic stress-strain curves of the tensile tests of the 3D printed material samples with different geometrical angles $\theta = 0°, 15°, 30°, 45°, 60°, 75°$.

When $\theta = 0°$, the metamaterial is reciprocal elastic ($\epsilon = 0$) where only one stress-strain curve was obtained when testing the material under the two setups ($A \to B$ and $B \to A$) (Fig. 1(d)). This material exhibited the same elasticity no matter what the direction of the material sample is with respect to the applied tensile stress direction. However, the other material samples with geometrical angles $\theta > 0$ showed a robust stress direction dependence, and exhibited asymmetric deformations and significant nonreciprocal elasticities ($\epsilon \neq 0$). Two different stress-strain curves were obtained when testing the material under the two experimental setups ($A \to B$ and $B \to A$). In the setup $A \to B$, the material showed a strong stiffness against the applied tensile stress while it appeared of a lower stiffness in the other setup ($B \to A$). The contrast in the material stiffnesses can be attributed to the geometrical asymmetry that has been achieved by tilling the struts. The nonreciprocal elasticity increases due to an increase in the geometrical angle $\theta$ (Fig. 1(c)).

It follows from Fig.1(d) that the asymmetry in the material deformation can be achieved using linear elastic materials. We tested a material sample, which was intentionally 3D printed of a filament with a higher elastic modulus, to give two different linear elastic stress-strain curves ($\theta = 30°$). Despite this material sample is linear elastic, it exhibited a significant nonreciprocal elasticity. These results indicate that breaking the asymmetry in the material deformation can be achieved by linear materials if a suitable microstructural-geometrical asymmetry is crafted into the material. In other words, the nonlinearity is not enough to achieve asymmetric deformation of the material or to realize mechanical nonreciprocity. Nonetheless, concurrently achieving nonlinearity and microstructural asymmetries promotes the mechanical nonreciprocity[2,8,10,17–19].

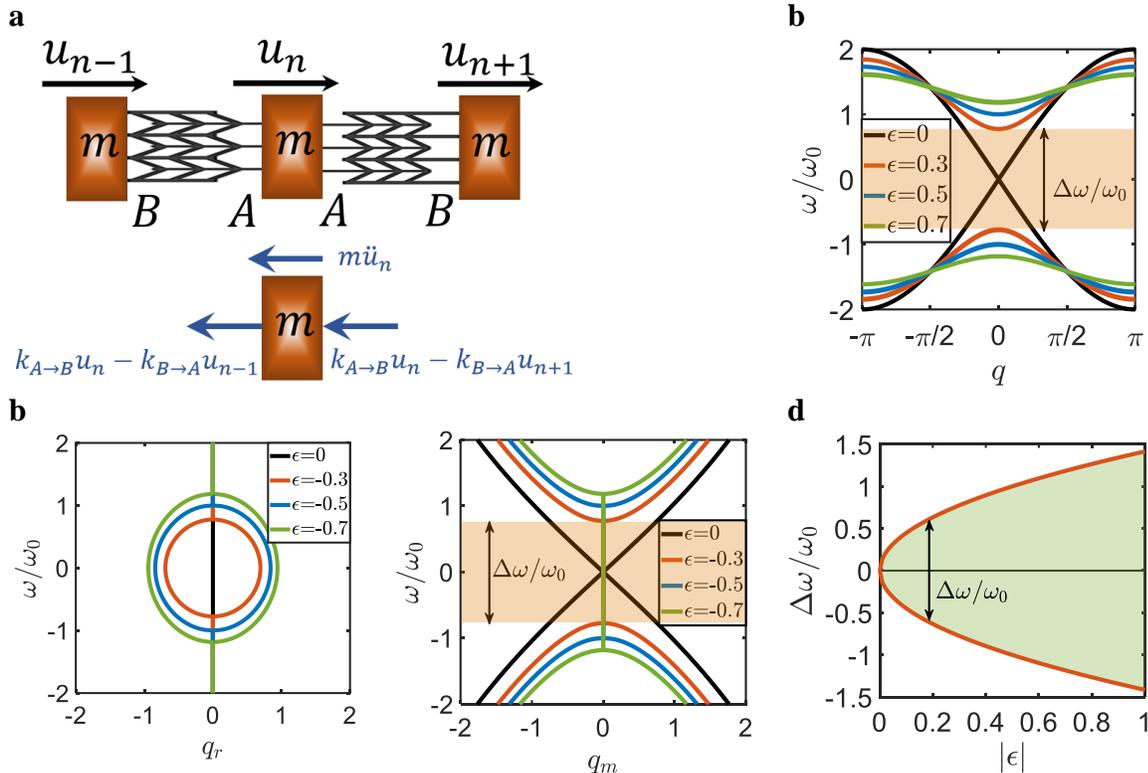

**Fig. 2: Topological Mechanics of Monoatomic Lattices with Nonreciprocal Elastic Springs. a**, A mass-spring system representing a monoatomic lattice ($m$ indicates the atomic mass). Nonreciprocal springs made of the developed metamaterial are considered such that the stiffness is different when the spring is stretched/compressed from two opposite ends ($k_{A \to B} \neq k_{B \to A}$). The free body diagram of the spring forces and the inertia forces is represented. **b**, Classical band structures $\omega(q)$ (nondimensional frequency ($\omega/\omega_0$) versus nondimensional wavenumber ($q$)) are obtained by implementing springs with positive nonreciprocal elasticity ($\epsilon > 0$). **c**, Complex band structures are obtained by implementing springs with negative nonreciprocal elasticity ($\epsilon < 0$) (real band structures $\omega(q_r)$ (left) and imaginary band structures $\omega(q_m)$ (right)). **d**, The band-gap ($\Delta\omega$) versus the nonreciprocal elasticity parameter ($\epsilon$).

By means of topological mechanics, we demonstrate that the trigger of the mechanical nonreciprocity is a "nonreciprocal elasticity". We investigated the topological mechanics of monoatomic and diatomic lattices. The conventional reciprocal springs of these lattices were replaced by nonreciprocal springs made of the proposed metamaterial (Figs.2(a) and 3(a)).

Implementing the proposed metamaterial gave the spring stiffness to a stretch/compression from one end is different than the spring stiffness to a stretch/compression from the other end, i.e., $k_{A \to B} \neq k_{B \to A}$ (Figs.2(a) and 3(a)). The nonreciprocal elasticity of these springs was defined by $\epsilon = 1 - (k_{B \to A} / k_{A \to B})$.

First, we studied the emerging topological properties of monoatomic lattices with nonreciprocal springs by investigating their classical and complex band structures (Figs. 2(b) and 2(c)). Gaps were observed in the band structures when nonreciprocal springs were implemented ($\epsilon \neq 0$) (Fig. 2(d)) (the band-gap vanishes if and only if reciprocal elastic springs are used ($\epsilon = 0$)). The existence of a vanishing gap when reciprocal springs are implemented (i.e., $\epsilon = 0$) indicates that the emerging topology of monoatomic lattices with nonreciprocal springs is non-trivial[2,20,21], and hence the trigger of the mechanical nonreciprocity is a nonreciprocal elasticity.

Then, we studied the topological properties of diatomic lattices with nonreciprocal elastic springs. Fig. 3(b) shows the band structures of the diatomic lattice for cases when $\epsilon = 0.5$, $0$, and $-0.5$. Band-gaps ($\Delta \omega$) are noticeably seen in the band structures as long as $\epsilon \neq 0$ (i.e., $\Delta \omega = \omega_A \left| \sqrt{2} - \sqrt{2 - 2\epsilon} \right|$). Whereas the band structures of the diatomic lattice obtained for positive and negative nonreciprocal elasticities, $\epsilon = 0.5$ and $-0.5$ would look similar, they are – in fact – different in the band-gap structure and the evolution of the eigenmodes. We observed "Eigenvalue Loci Veering" due to a change in the nonreciprocal elasticity (Fig.3(c)). The acoustic and optical frequencies approach each other and then veer apart as $\epsilon$ changes from positive to negative. The transition from "approaching" to "veering" of the acoustic and optical frequencies takes place exactly at $\epsilon = 0$ (where the gap is completely closed). This interesting observation indicates that the eigenvalue loci veering of diatomic lattices requires non-trivial topology. In other words, the eigenvalue loci veering of diatomic lattices cannot be achieved without closing the gap. The eigenvalue loci veering is mainly due to a rapid variation in the eigenvectors, which would result in either a mode inversion[22] (band inversion[23]) or a mode localization[24,25] (band localization). This is further investigated in Fig. 3(d), by observing the evolution of the acoustic and optic eigenvectors $\{1 \quad \Psi(q)\}^{\mathrm{T}}$.

It follows from Figs. 3(c) and 3(d) that the acoustic and optical frequencies and mode shapes are distinct when $0 < q < \pi$. The acoustic mode is in-phase where $\mathrm{Re}(\Psi(q)) > 0$, while the optical mode is out-of-phase where $\mathrm{Re}(\Psi(q)) < 0$. The eigenvalue loci veering takes place when $q = \pi$, and it causes mode (or band) localization without band inversion. The contours of the acoustic (optic) eigenmodes are confined to the right (left) of the complex plane (Fig. 3(d)). This indicates no winding about the origin and no band inversion. A closed contour indicates a band localization due to the eigenvalue loci veering where the vibration energy is inhibited to be localized at only one atom of the diatomic lattice (Fig. 3(d)). When $\epsilon = 0$, the acoustic and optic contours are closed, which indicates that the acoustic and optic bands have the same frequency and mode shape, and are localized when $q = \pi$, i.e., $\mathrm{Re}(\Psi(q)) = 0$ and $\mathrm{Im}(\Psi(q)) = 0$. When $\epsilon \neq 0$, the band localization occurs for either acoustic or optical bands depending on the sign of $\epsilon$. When $\epsilon > 0$, the optical contour is closed, which indicates band localization. On the other hand, the acoustic contour is open indicating no band localization where the vibration energy distributes between the two atoms. When $\epsilon < 0$, acoustic and optical bands switch roles, where acoustic bands are localized.

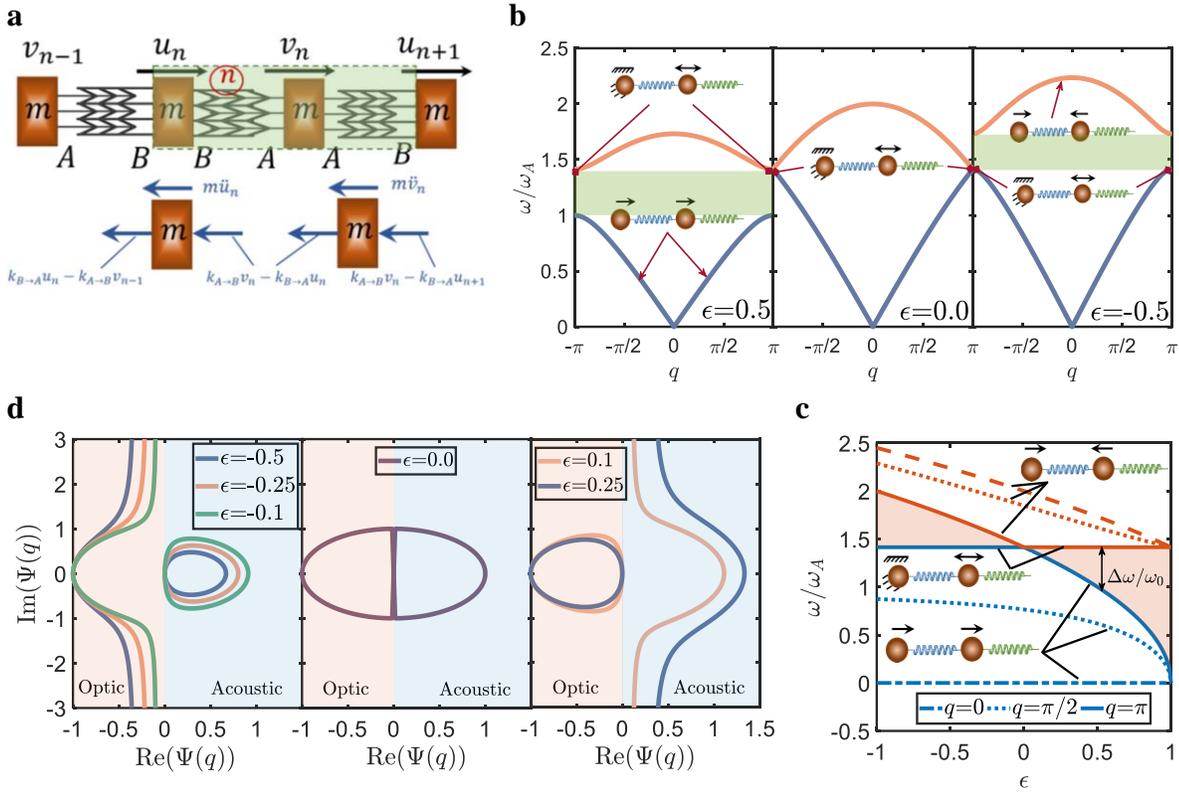

**Fig. 3: Topological Mechanics of Diatomic Lattices with Nonreciprocal Elastic Springs. a**, A diatomic lattice model. Nonreciprocal springs made of the developed metamaterial are considered such that the stiffness is different when the spring is stretched/compressed from two opposite ends ($k_{A \to B} \neq k_{B \to A}$). The free body diagram of the spring forces and the inertia forces is represented. **b**, The band structure $\omega(q)$ of the diatomic lattice for different values of the nonreciprocal elasticity parameter, $\epsilon = 0.5$ (left), $\epsilon = 0$ (middle), and $\epsilon = -0.5$ (right). **c**, The acoustic and optical frequencies at different wavenumbers, $q = 0, \pi/2$ and $\pi$, versus the nonreciprocal elasticity parameter $\epsilon$. The insets represent the Floquet–Bloch eigenmodes. **e**, Contour plots of the acoustic and optic eigenmodes $\Psi(q)$ in the complex plan for different values of the nonreciprocal elasticity parameter $\epsilon$

These observed differences in the mode shapes of the acoustic and optical bands are associated with a band-gap formation. This demonstrates that the formation of band-gaps in the band structure of diatomic lattices with nonreciprocal springs requires a localization of either the acoustic energy or the optical energy at only one atom. However, no band-gaps are formed if both acoustic and optical energies are localized. Generally speaking, the formation of the band-gap of a topological mechanical system would require a confinement of the vibration energy to one of the bands, which can be achieved if a proper nonreciprocal elasticity is implemented.

In this paper, we developed metamaterials with nonreciprocal linear and nonlinear elasticities. We demonstrated that the trigger of the nonreciprocity of static mechanical systems is not essentially the material to be nonlinear but a nonreciprocal elasticity. We achieved nontrivial topological static mechanical systems utilizing nonreciprocal linear and nonlinear elastic materials. We expect that the developed nonreciprocal elastic metamaterial can impart effective nonreciprocity to static and dynamic mechanical systems.

**Acknowledgments**

**Funding:** This research is supported by Abu Dhabi University (Grants 19300474 and 19300475).

**Competing interests:** The author has no competing interests to disclose.

**Data and materials availability:** The data that support the findings of this study are included in the article.